\documentclass[12pt,axodraw]{article}
\usepackage{amsmath,url} 
\usepackage{axodraw}
\usepackage{graphicx}
\def\ltap{\raisebox{-.6ex}{\rlap{$\,\sim\,$}} \raisebox{.4ex}{$\,<\,$}} 
\def\gtap{\raisebox{-.6ex}{\rlap{$\,\sim\,$}} \raisebox{.4ex}{$\,>\,$}}

\newcommand\as{\alpha_{\mathrm{S}}} 
\newcommand\f[2]{\frac{#1}{#2}} 
 
\def\to{\rightarrow}
 
\def\nn{\nonumber}

\def\ms{${\overline {\rm MS}}$}
\def\mH{m_H}
\def\mb{m_b}
\def\mt{m_t}
\def\pT{p_T}
\def\tL{{\widetilde L}}

\begin{document} 
\begin{titlepage}
\renewcommand{\thefootnote}{\fnsymbol{footnote}}
\begin{flushright}
ZU-TH 10/13
\end{flushright}
\vspace*{2cm}

\begin{center}
{\Large \bf Heavy-quark mass effects\\[0.2cm] in Higgs boson production at the LHC}
\end{center}

\par \vspace{2mm}
\begin{center}
{\bf Massimiliano Grazzini\footnote{On leave of absence from INFN, Sezione di Firenze, Sesto Fiorentino, Florence, Italy.}}
and
{\bf Hayk Sargsyan}

\vspace{5mm}

Institut f\"ur Theoretische Physik, Universit\"at Z\"urich, CH-8057 Z\"urich, Switzerland

\vspace{5mm}

\end{center}

\par \vspace{2mm}
\begin{center} {\large \bf Abstract} \end{center}
\begin{quote}
\pretolerance 10000

We study the impact of heavy-quark masses in Higgs boson production through gluon fusion at the LHC. We extend previous computations of the fully differential cross section and of the transverse momentum spectrum of the Higgs boson by taking into account the finite top- and bottom-quark masses up to ${\cal O}(\as^3)$. We also discuss the issues arising when the heavy-quark mass is much smaller than the Higgs mass. Our results are implemented in updated versions of the {\tt HNNLO} and {\tt HRes} numerical programs.

\end{quote}

\vspace*{\fill}
\begin{flushleft}
June 2013

\end{flushleft}
\end{titlepage}

\setcounter{footnote}{1}
\renewcommand{\thefootnote}{\fnsymbol{footnote}}

\section{Introduction}
\label{sec:intro}

The discovery of a new boson with mass $\mH\sim 125$ GeV by the ATLAS and CMS experiments at the LHC \cite{Aad:2012tfa,Chatrchyan:2012ufa} has lead, if at all possible, to a renewed interest in Higgs physics.
Although the discovery is independent on the details of the theoretical modelling of the signal, to establish the extent to which the new resonance is consistent with the long sought Higgs boson requires accurate theoretical predictions for the Higgs production cross section and the associated
distributions.

The main production mechanism of the Standard Model (SM) Higgs boson at hadron colliders is gluon fusion through a heavy-quark loop \cite{Georgi:1977gs}. At the LHC the $gg\rightarrow H$ cross section is typically one order of magnitude larger than the cross section in the other channels for a wide range of the Higgs boson masses.
It is thus essential to have the theoretical predictions for $gg\to H$ under very good control, and this implies
an accurate evaluation of radiative corrections.

As the Higgs coupling to quarks is proportional to their masses, the main contributions arise from the top and bottom quarks. The calculation of radiative corrections is simplified in the so-called large-$\mt$ approximation, in which the fermionic loop is replaced by an effective vertex and, thus, the number of loops is reduced by one. The QCD radiative corrections to this process at next-to-leading order (NLO) are known both in the large-$\mt$ limit \cite{Dawson:1990zj} and by keeping the exact dependence on the masses of the top and bottom quarks \cite{Djouadi:1991tka,Graudenz:1992pv,Spira:1995rr}. It turns out that the NLO corrections increase the cross section by 80-100$\%$ at the LHC \cite{Dawson:1990zj,Djouadi:1991tka}. The next-to-next-to-leading (NNLO) corrections are known only in the large-$\mt$ limit and increase the cross section by about 25 $\%$ \cite{Harlander:2002wh, Anastasiou:2002yz, Ravindran:2003um}.
The NNLO corrections have been successfully implemented in two independent
fully-exclusive numerical programs {\tt FehiPro} \cite{Anastasiou:2004xq,Anastasiou:2007mz,Anastasiou:2009kn} and {\tt HNNLO} \cite{Catani:2007vq,Grazzini:2008tf}, which allow the user to compute the Higgs production cross section
by applying arbitrary kinematical cuts on the Higgs decay products and the associated jet activity.

At hadron colliders the production of an (on shell) Higgs boson is characterized by
its transverse momentum $\pT$ and
rapidity $y$. The rapidity distribution is essentially driven by the parton distribution functions
of the partons in the colliding hadrons, and it is mildly sensitive to radiative corrections.
By contrast, the $\pT$ distribution is sensitive to multiple emissions from the initial state partons and
its detailed knowledge is very important in the experimental analyses.
A measurement of the Higgs $\pT$ spectrum at the LHC
is to be expected in the near future.

When $\pT\sim \mH$ the QCD radiative corrections to the transverse momentum cross section $d\sigma/d\pT$ can be evaluated through the standard fixed-order expansion. When $\pT\ll \mH$ the convergence of the perturbative expansion is spoiled by the presence of
large logarithmic terms. To obtain reliable perturbative predictions over the whole range of transverse momenta, such terms must be resummed to all orders, and the result has to be consistently matched
to the standard fixed-order result valid at $\pT\sim \mH$.
A computation of the resummed $\pT$ spectrum up to next-to-next-to-leading logarithmic accuracy \cite{deFlorian:2001zd,Becher:2010tm,Catani:2011kr}, matched to the ${\cal O}(\as^4)$ result valid at large $\pT$  \cite{deFlorian:1999zd,Glosser:2002gm,Ravindran:2002dc} is implemented in the numerical program {\tt HqT} \cite{Bozzi:2003jy,Bozzi:2005wk,deFlorian:2011xf}.
An extension of this program, including the decay of the Higgs boson in the $\gamma\gamma$, $WW$ and $ZZ$ final states is implemented in the code {\tt HRes} \cite{deFlorian:2012mx}.
Both these calculations are performed by using the large-$\mt$ approximation.

The purpose of the present paper is twofold. We first document the inclusion of heavy-quark mass effects
up to NLO in the fully exclusive computation of Refs.~\cite{Catani:2007vq,Grazzini:2008tf}.
We then address the implementation of mass effects in the resummed $\pT$ spectrum. As far as top-mass effects are concerned, since $\mt\sim \mH$, the computation of the Higgs spectrum is still a {\em two scale} problem, and the implementation does not lead to substantial complications.
The inclusion of bottom-mass effects is instead more difficult. Since $\mb\ll\mH$, the computation of the $\pT$ spectrum becomes a {\em three scale} problem, whose solution beyond the fixed order is by far non trivial.
We propose a simple treatment for this issue and we present new results for the  resummed $\pT$ spectrum and 
the ensuing uncertainties. 

The paper is organized as follows. In Sec.~2 we discuss the implementation of the exact top- and bottom-mass dependence in the fully exclusive NNLO computation of Refs.~\cite{Catani:2007vq,Grazzini:2008tf} and present some numerical results. In Sec.~3 we discuss the implementation of the mass effects in the resummed $\pT$ spectrum.
The issues arising when the heavy-quark mass is much smaller than the Higgs mass are discussed in Sec.~3.1, where we
also present our treatment of the bottom quark in the resummed $\pT$ spectrum. In Sec.~3.2 we present our numerical results. In Sec.~4 we draw our conclusions.

\section{Mass effects at fixed order}

\label{sec:HNNLO}

In the following we briefly recall the subtraction formalism developed in Ref.~\cite{Catani:2007vq} to deal with the soft and collinear divergences appearing in real and virtual QCD corrections at NLO and NNLO, and we discuss how the NNLO calculation of Refs.~\cite{Catani:2007vq,Grazzini:2008tf} can be extended to include heavy-quark masses up to NLO.

We briefly introduce the theoretical framework and our notation.
We consider the inclusive hard scattering process
\begin{equation}
h_1+h_2 \rightarrow H+X,
\end{equation}
where the collision of the two hadrons produces the Higgs boson $H$ accompanied by an arbitrary and undetected final state $X$.
The LO partonic subprocess is the gluon fusion mechanism $gg \to H$.
We use the narrow-width approximation and we treat the Higgs boson as an
on-shell particle with mass $\mH$.
We use parton
densities as defined in the \ms\
factorization scheme, and we denote by $\as(\mu_R^2)$ the QCD running coupling 
at the renormalization scale $\mu_R$ in the \ms\
renormalization scheme.

According to the formalism of Ref.~\cite{Catani:2007vq} the {\em fully differential} cross section at (N)NLO can be schematically written as
\begin{equation}
d\sigma_{(N)NLO}={\cal H}^{(N)NLO}\otimes d\sigma_{LO}+\left[d\sigma^{H+{\rm jet(s)}}_{(N)LO}-d\sigma^{CT}_{(N)LO}\right].
\label{eq:subnnlo}
\end{equation}
The first term on the right-hand side of Eq. (\ref{eq:subnnlo}) contains the LO cross section $d\sigma_{LO}$ suitably convoluted with a perturbatively computable hard-collinear function ${\cal H}^{(N)NLO}$. The second term of the right hand side of Eq.~(\ref{eq:subnnlo}) contains the (N)LO cross section for
the $H+{\rm jet(s)}$ process, $d\sigma^{H+{\rm jet(s)}}_{(N)LO}$. This cross section is finite as soon as the transverse momentum $\pT$ of the Higgs boson is non vanishing, and can be obtained with any available method to
perform NLO QCD computations. The first NLO calculation of $d\sigma^{H+{\rm jet(s)}}$ was
presented in Ref.~\cite{deFlorian:1999zd}, and was performed by using
the FKS version of the subtraction formalism \cite{Frixione:1995ms,Frixione:1997np}. 
In our numerical implementation we compute $d\sigma^{H+{\rm jet(s)}}$ with the dipole subtraction method \cite{Catani:1996jh,Catani:1996vz} as implemented in the MCFM numerical program \cite{MCFMweb}. The singularity at $\pT\to 0$ is subtracted by using the counterterm $d\sigma^{CT}_{N(LO)}$, which is computed as \cite{Catani:2007vq}
\begin{equation}
d\sigma^{CT}_{N(LO)}=d\sigma_{LO}\otimes \Sigma(\pT/\mH)d^2{\bf p}_T\, .
\label{eq:ct}
\end{equation}
The function $\Sigma^H(\pT/\mH)$ depends only on the channel in which the process occurs at Born level (gluon fusion in the present case) and it embodies the singular behaviour of the Higgs $\pT$ distribution as $\pT\to 0$ \cite{Bozzi:2005wk}.

As far as the first term in Eq.~(\ref{eq:subnnlo}) is concerned, the exact top and bottom mass dependence up to NLO
can be implemented by replacing the Born cross section $d\sigma_{LO}$ evaluated in the large-$\mt$ limit
with the cross section with the exact dependence on the top and bottom masses \cite{Georgi:1977gs}, and by computing the exact expression of the coefficient ${\cal H}^{NLO}$.

The coefficient ${\cal H}^{NLO}$ is
\begin{equation}
{\cal H}^{NLO}_{ab}(z_1,z_2,\as)=\delta_{ga}\delta_{gb}\delta(1-z_1)\delta(1-z_2)+
\left(\f{\as}{\pi}\right){\cal H}_{ab}^{(1)}(z_1,z_2)\, ,
\end{equation}
where ${\cal H}_{gg}^{(1)}$ contains the information on the virtual correction to the LO subprocess and is given by
\begin{equation}
{\cal H}^{(1)}_{gg}(z_1,z_2)=\delta(1-z_1)\delta(1-z_2)
\left(C_A\frac{\pi^2}{6}+\frac{1}{2}\mathcal{A}\right)\, ,
\label{Hgg1}
\end{equation}
whereas the off diagonal coefficients read
\begin{equation}
{\cal H}^{(1)}_{gq}(z_1,z_2)=\frac{1}{2}C_{F}z_2\, \delta(1-z_1)~~~~~~~
{\cal H}^{(1)}_{qg}(z_1,z_2)=\frac{1}{2}C_{F}z_1\, \delta(1-z_2)~~~~~~~
{\cal H}^{(1)}_{q\bar{q}}(z_1,z_2)=0\, .
\label{Hgqqq}
\end{equation}
The function $\mathcal{A}$ in Eq. (\ref{Hgg1}) denotes the finite part of the virtual correction to $gg\to H$ defined
according to the conventions of Ref.~\cite{deFlorian:2001zd}, and, in the large-$m_t$ approximation it reads
\begin{equation}
\mathcal{A}=5C_A+\frac{2}{3}C_A\pi^2-3C_F\equiv 11+2\pi^2.
\label{AA}
\end{equation}
To obtain the exact form of ${\cal H}^{NLO}$ it is enough to replace the coefficient ${\cal A}$ in Eq.~(\ref{AA})
with the corresponding function with the exact dependence on the masses of the top and bottom quarks.
The result can be found in Eq.~(B.2) of Ref.~\cite{Spira:1995rr} in terms of one-dimensional integrals, or as a fully analytic expression in Eq.~(3.5) of Ref.~\cite{Harlander:2005rq} and Eq.~(27) of Ref.~\cite{Aglietti:2006tp}, both in terms of harmonic polylogarithms. In our numerical implementation we use the result of Ref.~\cite{Spira:1995rr}, as implemented
in the {\tt HIGLU} numerical program.

We finally discuss the implementation of the heavy-quark masses in the second contribution on the right-hand side of Eq.~(\ref{eq:subnnlo}). The cross section $d\sigma_{N(LO)}^{H+{\rm jet(s)}}$ can be evaluated by replacing the ${\cal O}(\as^3)$ $H$+3 parton matrix element with its exact expression as a function of the heavy-quark masses \cite{Ellis:1987xu}.
Correspondingly, the ${\cal O}(\as^3)$ contribution to the subtraction counterterm $d\sigma^{CT}_{(N)LO}$ must be evaluated by using the exact expression for the Born cross section $d{\sigma}_{LO}$ in Eq.~(\ref{eq:ct}), so as to cancel the singular behaviour
at small $\pT$.

The procedure defined above allows us to compute the fully differential Higgs production cross section up to NLO according to Eq.~(\ref{eq:subnnlo}). The NNLO matrix elements \cite{Kauffman:1996ix,Schmidt:1997wr,Harlander:2000mg}
are known only in the large-$\mt$ approximation\footnote{Corrections to the large-$\mt$ approximation at NNLO have been considered
in Refs.~\cite{Marzani:2008az,Harlander:2009bw,Harlander:2009mq,Harlander:2009my,Pak:2009bx,Pak:2009dg,Harlander:2012hf}.}.
As a consequence, at ${\cal O}(\as^4)$ we include only the top-quark contribution, evaluated in the large-$\mt$ approximation, and we normalize it with the exact $\mt$-dependent Born cross section, $\sigma_{LO}(\mt)$.
More precisely, we multiply the ${\cal O}(\as^4)$ contributions by the ratio $\sigma_{LO}(\mt)/\sigma_{LO}(\mt\to\infty)$.

\subsection{Numerical results}

We have implemented the exact heavy-quark mass dependence in a new version of the numerical code {\tt HNNLO}. The program {\tt HNNLO} is a parton level event generator that allows the user to compute the Higgs production cross section and the associated distributions up to NNLO in QCD perturbation theory, and to apply arbitrary infrared-safe cuts on the Higgs decay products and the recoiling QCD radiation. The program includes the $H\to\gamma\gamma$, $H\to WW\to l\nu l\nu$ and $H\to ZZ\to 4l$ decay modes.

In the following, we present only a limited sample of the numerical results that can be obtained with our program.
We consider Higgs boson production in $pp$ collisions at $\sqrt{s}=8$ TeV and
we use the MSTW2008 sets of parton distributions \cite{Martin:2009iq}, with densities and $\as$ evaluated at each corresponding order (i.e., we use $(n+1)$-loop $\as$ at N$^n$LO). Unless stated otherwise, we set the renormalization and factorization scales to the Higgs boson mass, $\mu_R=\mu_F=\mH$, and we set $\mt=172.5$ GeV and $\mb=4.75$ GeV.

The first quantity that is important to test with the modified program is the inclusive cross section.
In Table~1 we study the impact of heavy-quark masses at NLO.
We report the NLO cross sections evaluated with the exact top and bottom mass dependence, normalized to the NLO result in the large-$\mt$ limit.

\begin{table}[ht]
\begin{center}
\begin{tabular}{|c| c| c|}
\hline\hline
$m_H(\mathrm{GeV})$ & $\frac{\sigma_{NLO}(m_t)}{\sigma_{NLO}(m_t\rightarrow\infty)}$ & $\frac{\sigma_{NLO}(m_t,m_b)}{\sigma_{NLO}(m_t\rightarrow\infty)}$ \\ [0.5ex]
\hline
125 & 1.061 & 0.988 \\
\hline
150 & 1.093 & 1.028 \\
\hline
200 & 1.185 & 1.134 \\
\hline
\end{tabular}
\end{center}
\label{table1}
\caption{Impact of the heavy-quark masses on the inclusive NLO cross sections. All results are normalized to the $m_t\rightarrow\infty$ result.}
\end{table}

From Table~1 we see that the mass effects change the cross section at the few percent level, and that the bottom contribution decreases the cross section by a few percent. This effect is well known, and it is due to the negative interference with the top-quark
contribution.
We have compared our results with those obtained with the numerical program {\tt HIGLU} \cite{Djouadi:1991tka,Spira:1995rr} and found very good agreement.

We now move to consider the impact of mass effects on
the $\pT$ cross section. Such effects have been studied at NLO in earlier
works \cite{Baur:1989cm,Langenegger:2006wu,Keung:2009bs,Anastasiou:2009kn,Bagnaschi:2011tu,Mantler:2012bj}.

\begin{figure}[htb]
\begin{center}
\includegraphics[scale=0.55]{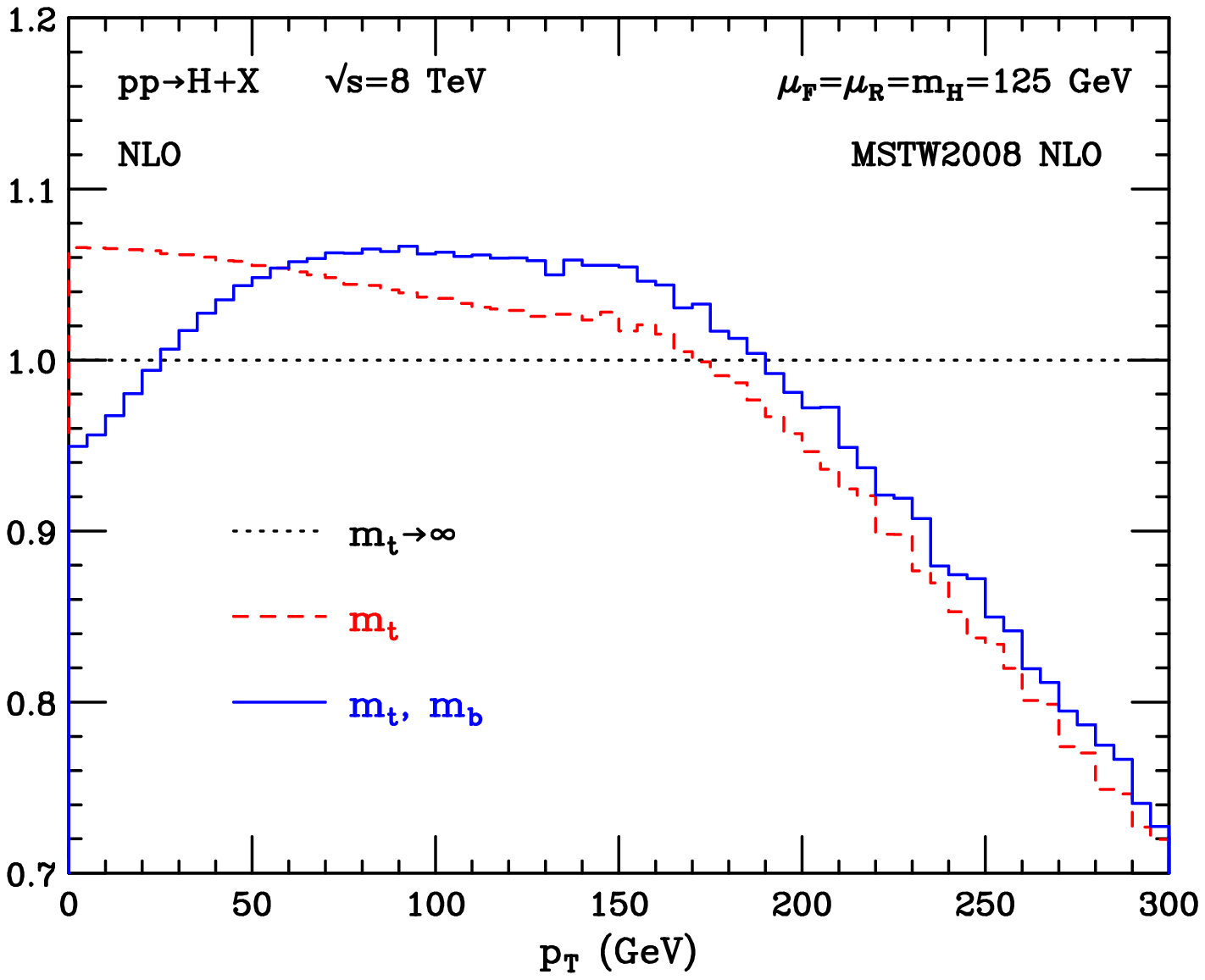}
\includegraphics[scale=0.55]{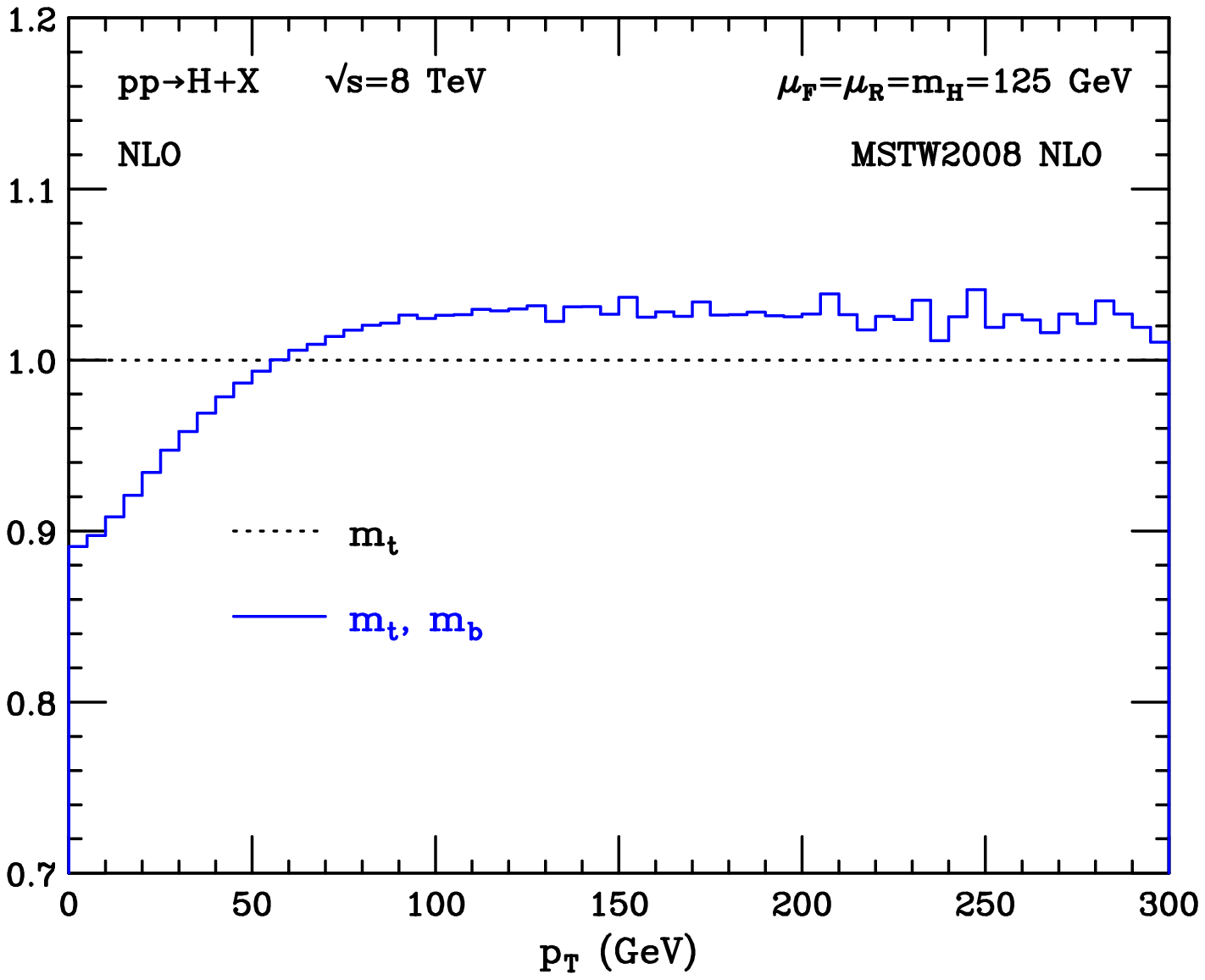}\\
\caption{Transverse momentum distribution for a SM Higgs with $m_H=125$ GeV computed at NLO. Left: result normalized to the large-$\mt$ approximation. Right: normalized to the $m_t$-dependent result.}
\label{fig:ptdistmtmb}
\end{center}
\end{figure}

In Fig.~\ref{fig:ptdistmtmb} (left panel) we plot the $\pT$ spectrum of the Higgs boson at NLO with full dependence on the masses of the top and bottom quarks and
we compare it with the corresponding result in which only the top-quark contribution is considered. Both results are normalized to the result obtained in the large-$\mt$ limit. To better emphasize the impact of the bottom quark, in the right panel of Fig.~\ref{fig:ptdistmtmb} we show the full NLO result normalized to the result obtained neglecting the bottom quark.

We see that, when only the top contribution is considered, the cross section
at low $\pT$ is larger than the corresponding cross section in the large-$\mt$ limit. In this region the recoiling parton is soft and/or collinear, and the
differential cross section factorizes into a universal factor times the Born level contribution. The limit of the solid and dashed histograms in the left panel of Fig.~\ref{fig:ptdistmtmb} thus correspond to the ratios $\sigma_{LO}(\mt,\mb)/\sigma_{LO}(\mt\to \infty)=0.949$ and $\sigma_{LO}(\mt)/\sigma_{LO}(\mt\to \infty)=1.066$, respectively.

The results in Fig.~\ref{fig:ptdistmtmb} show that the impact of the bottom quark is important, especially in the low-$\pT$ region, since it substantially deforms the shape of the spectrum. At large $\pT$ values, the impact of the bottom quark becomes small and the differential
cross section quickly departs from its value in the large-$\mt$ limit.
This is a well known feature of the large-$\mt$ approximation: at large $\pT$ the parton recoiling against the Higgs boson is sensitive to the heavy-quark loop, and the large-$\mt$ approximation breaks down.

Another feature that is evident from Fig.~\ref{fig:ptdistmtmb} is that the qualitative behaviour of the results is rather different. When considering the NLO result with only the top quark included, in a wide region of transverse momenta the shape of the spectrum is rather stable and in rough agreement with what is obtained in the large-$\mt$ approximation. This is not the case when the bottom contribution is included: the shape of the spectrum quickly changes in the small- and intermediate-$\pT$ region and the spectrum becomes harder.
We will come back to this point in Sec.~\ref{sec:bottom}.

%%%%%%%%%%%%%%%%%%%%%%%%%%%%%%%%%%%%%%%%%%%%%%%%%
%%%%%%%%%%%%%%%%%%%%%%%%%%%%%%%%%%%%%%%%%%%%%%%%%

The mass effects in differential NLO distributions
were previously discussed in Ref.~\cite{Anastasiou:2009kn}. 
We have compared our results with those of Ref.~\cite{Anastasiou:2009kn} and found agreement.

\section{Mass effects in the resummed $\pT$ spectrum}

\label{sec:resu}

As we mentioned in the previous Section, the fixed order transverse momentum distribution diverges
at small $p_T$. In order to obtain a reliable behaviour in this region the divergent contributions
need to be resummed to all orders in perturbation theory.  Once the resummation has been carried out,
the result has to be properly matched to the standard fixed order result, so as to obtain a prediction
which is valid in the entire range of transverse momenta.
In this Section we briefly recall the resummation procedure of Refs.~\cite{Bozzi:2005wk,Bozzi:2007pn}
and we comment on the inclusion of mass effects.

The QCD expression of the Higgs boson doubly differential
cross section is
\begin{align}
\label{dcross}
\f{d\sigma}{dy\,d \pT^2}(y,\pT,\mH,s)&=\sum_{a,b}
\int_0^1 dx_1 \,\int_0^1 dx_2 \,f_{a/h_1}(x_1,\mu_F^2)
\,f_{b/h_2}(x_2,\mu_F^2)\nn\\
&\times\f{d{\hat \sigma}_{ab}}{d{\hat y}\,d \pT^2}({\hat y},\pT, \mH,{\hat s}; \as(\mu_R^2),\mu_R^2,\mu_F^2) 
\;,
\end{align}
where $f_{a/h_i}(x,\mu_F^2)$ ($a=q_f,{\bar q_f},g$) are the parton densities of 
the colliding hadrons ($h_1$ and $h_2$) at the factorization scale $\mu_F$,
and $d{\hat \sigma}_{ab}$ are the
partonic cross sections. The centre--of--mass energy of the two colliding
hadrons is denoted by $s$, and ${\hat s}$ is the partonic centre--of--mass
energy.
The rapidity, $\hat y$, and the centre-of-mass energy,
${\hat s}$, of the partonic subprocess are related to
the corresponding hadronic variables $y$ and $s$ as:
\begin{equation}
\label{kin}
{\hat y} = y - \frac{1}{2} \ln\frac{x_1}{x_2} \;, \quad 
\quad {\hat s}=x_1x_2s \;\;.
\end{equation}
The partonic cross section $d{\hat \sigma}_{ab}$ is computable in QCD 
perturbation theory but its series expansion in $\as$ contains  
the logarithmically-enhanced terms, $(\as^n/p_T^2)\, \ln^m (m_H^2/p_T^2)$,
that we want to resum.

To this purpose, the partonic cross section is rewritten as the sum of two terms,
\begin{equation}
\label{resfin}
\f{d{\hat \sigma}_{a_1a_2}}{d{\hat y} \,dp_T^2} =
\f{d{\hat \sigma}_{a_1a_2}^{(\rm res.)}}{d{\hat y} \,dp_T^2}
+\f{d{\hat \sigma}_{a_1a_2}^{(\rm fin.)}}{d{\hat y} \,dp_T^2} \;\;.
\end{equation}
The logarithmically-enhanced contributions are embodied in the 
`resummed' component $d{\hat \sigma}_{a_1a_2}^{(\rm res.)}$.
The `finite' component $d{\hat \sigma}_{a_1a_2}^{(\rm fin.)}$
is free of such contributions, and it
can be computed by 
a truncation 
of the perturbative series at a given fixed order.
In particular we compute $d{\hat \sigma}_{a_1a_2}^{(\rm fin.)}$
starting from $\left[d{\hat \sigma}_{a_1a_2}\right]_{\rm f.o.}$,
the usual perturbative series truncated at a given fixed order in $\as$,
and we subtract the perturbative truncation of the resummed component at
the same order:
\begin{equation}
\left[\f{d{\hat \sigma}_{a_1a_2}^{(\rm fin.)}}{d{\hat y} \,dp_T^2}\right]_{\rm f.o.}=\Bigg[\f{d{\hat \sigma}_{a_1a_2}}{d{\hat y} \,dp_T^2}\Bigg]_{\rm f.o.}-
\left[\f{d{\hat \sigma}_{a_1a_2}^{(\rm res.)}}{d{\hat y} \,dp_T^2}\right]_{\rm f.o.}\, .
\end{equation}

The resummed component of the partonic cross section 
is obtained by working in impact parameter $b$ space
\begin{equation}
\label{resum}
\!\!\! \f{d{\hat \sigma}_{a_1a_2}^{(\rm res.)}}{d{\hat y} \,dp_T^2}({\hat y}, 
p_T,m_H,{\hat s};
\as)
= \f{m^2_H}{\hat s} \;
\int_0^\infty db \; \f{b}{2} \;J_0(b p_T) 
\;{\cal W}_{a_1a_2}({\hat y},b,m_H,{\hat s};\as)\;,
\end{equation}
where $J_0(x)$ is the 0th-order Bessel function, and the
factor ${\cal W}$
embodies the all-order dependence on 
the large logarithms $\ln (m_H^2b^2)$ at large $b$, which correspond to
$\ln (m^2_H/p_T^2)$ terms in $p_T$ space.

In the case of the $p_T$ cross section integrated over the rapidity,
it is convenient to define \cite{Catani:2000vq,Bozzi:2005wk}
the $N$-moments ${\cal W}_N$ of ${\cal W}$
with respect to $z=m^2_H/{\hat s}$ at fixed $m_H$. In the case
in which the dependence on the rapidity is taken into account, it is useful
to consider `double' $(N_1,N_2)$-moments
with respect to the two variables
$z_1=e^{+{\hat y}} m_H/{\sqrt{\hat s}}$ and 
$z_2=e^{-{\hat y}} m_H/{\sqrt{\hat s}}$ at fixed $m_H$
(note that $0< z_i <1$).
We thus introduce ${\cal W}^{(N_1,N_2)}$ as follows \cite{Bozzi:2007pn}:
\begin{equation}
\label{wnnudef}
{\cal W}_{a_1a_2}^{(N_1,N_2)}(b,m_H;\as) =
\int_0^1 dz_1 \,z_1^{N_1-1} \; \int_0^1 dz_2 \,z_2^{N_2-1} \;\, 
{\cal W}_{a_1a_2}({\hat y},b,m_H,
{\hat s};\as)
\;.
\end{equation}
More generally, for any function $h(y;z)$ with $|y|<-\ln \sqrt{z}$ and $0<z<1$
we define $(N_1,N_2)$ Mellin moments as
\begin{equation}
\label{melldef}
h^{(N_1,N_2)} \equiv
\int_0^1 dz_1 \,z_1^{N_1-1} \int_0^1 dz_2 \,z_2^{N_2-1}
\; h(y;z) \;, \quad {\rm where:} \; y=\frac{1}{2} \ln \frac{z_1}{z_2}\;,
\; z=z_1z_2 \;.
\end{equation}
By taking $(N_1,N_2)$ moments the QCD factorization formula (\ref{dcross}) becomes
\begin{equation}
\label{n12fact}
d\sigma^{(N_1,N_2)} = \sum_{a_1,a_2} \;f_{a_1/h_1, N_1+1} \;f_{a_2/h_2,N_2+1}
\;d{\hat \sigma}_{a_1a_2}^{(N_1,N_2)} \;,
\end{equation}
where $f_{a/h, N}= \int_0^1 dx \,x^{N-1} f_{a/h}(x)$ are the standard
Mellin moments of the parton distributions.

By using double Mellin moments the resummation structure of the
logarithmic terms in ${\cal W}_{a_1a_2}^{(N_1,N_2)}$
can be organized in exponential form as follows:
\begin{equation}
\label{wtilde}
{\cal W}^{(N_1,N_2)}(b,m_H;\as) =
\sigma_{LO}(\as,m_H)\,
{\cal H}^{(N_1,N_2)}(m_H,\as;m_H^2/Q^2)\,
\exp\{{\cal G}^{(N_1,N_2)}(\as,{\widetilde L}_Q;m_H^2/Q^2)\}
\;,
\end{equation}
where 
\begin{equation}
\label{logdef}
{\widetilde L}_Q= \ln\left(\f{Q^2\,b^2}{b_0^2} + 1\right) \;\;,
\end{equation}
$b_0=2e^{-\gamma_E}$ 
($\gamma_E=0.5772\dots$ is the Euler number) and, to simplify the notation,
the dependence on the flavour indices has been understood.
The scale $Q$ in Eq.~(\ref{logdef}), named resummation scale, parametrizes the arbitrariness in the resummation procedure. Its role is analogous
to the role played by the renormalization (factorization) scale in the context
of the renormalization (factorization) procedure. The resummed
cross section does not depend on $Q$ when evaluated at all perturbative orders,
but its dependence on $Q$ appears after truncation of the resummed expression at a given logarithmic accuracy.

The function ${\cal H}^{(N_1,N_2)}$ corresponds to the double Mellin transform of the function ${\cal H}$
introduced in Sect.~\ref{sec:HNNLO}. More precisely, the coefficients ${\cal H}(z_1,z_2,\as)$ correspond to
the Mellin inversion of ${\cal H}^{(N_1,N_2)}$ evaluated at $Q=\mH$.

The form factor $\exp\{{\cal G}\}$
includes the complete dependence on $b$ and,
in particular, it contains 
all the terms that order-by-order in $\as$ are logarithmically
divergent when $b \to \infty$. The functional dependence on $b$ is 
expressed through 
the large logarithmic terms $\as^n {\widetilde L}_Q^m$
with $1\leq m \leq 2n$. 
 
Note that we use the logarithmic variable ${\widetilde L}_Q$ 
(see Eq.~(\ref{logdef}))
to organize the resummation of the large logarithms $\ln (Q^2b^2)$.
In the region in which $Qb \gg 1$ we have
${\widetilde L}_Q \sim \ln (Q^2b^2)$ and the use of the variable 
${\widetilde L}_Q$ is fully legitimate to arbitrary logarithmic accuracy.
When $Qb \ll 1$, we have $\tL_Q \to 0$ 
and $\exp \{{\cal G}(\as, \tL_Q)\} \to 1$.
Therefore, the use of ${\widetilde L}_Q$ reduces the
effect produced by the resummed contributions
in the small-$b$ region (i.e., at large and intermediate values of $p_T$), 
where the large-$b$ resummation approach is not justified.
In particular, 
setting $b=0$ (which corresponds to integrate over the entire $p_T$ range)
we have $\exp \{{\cal G}(\as, \tL_Q)\} = 1$: this property
implies \cite{Bozzi:2005wk}
a unitarity constraint on the total cross section;
transverse-momentum resummation acts on the shape of the
$\pT$ distribution of the Higgs boson
without affecting its total production rate.

The formalism briefly recalled above defines 
a systematic expansion \cite{Bozzi:2005wk} of Eq.~(\ref{resfin})
whose orders are denoted as NLL+NLO, NNLL+NNLO and so forth.
In this notation the first label (NLL, NNLL, $\dots$) refers to the logarithmic accuracy at 
small $p_T$ and the second label (NLO, NNLO, $\dots$) refers to the customary 
perturbative order for the inclusive cross section.
More precisely, at NLL+NLO accuracy we use the NLL expression for the form factor
$\exp\{\cal{G}\}$, include the coefficient ${\cal H}^{(1)}$,
and expand the finite component to ${\cal O}(\as^3)$.
At NNLL+NNLO accuracy we use the NNLL expression for the form factor $\exp\{\cal{G}\}$,
include the coefficient ${\cal H}^{(2)}$ \cite{Catani:2011kr}, and truncate the finite component to ${\cal O}(\as^4)$.

We point out
that the NNLL+NNLO (NLL+NLO) result includes the {\em full} NNLO (NLO)
perturbative
contribution, supplemented with the resummation of the
logarithmically enhanced terms in the small-$p_T$ region at (N)NLL.

Having recalled the resummation formalism, we can now briefly comment on the
implementation of the exact top- and bottom-mass dependence in the resummed spectrum.
We note that the resummed cross section is proportional to the Born term (see Eq.~(\ref{wtilde})),
whose heavy-quark mass dependence can be implemented in a straightforward manner.
At NLL+NLO accuracy the additional mass dependence is embodied
in the coefficient ${\cal H}^{(1)}_{gg}$ in Eq.~(\ref{Hgg1})
whose mass dependence enters through the finite part of the virtual corrections $\mathcal{A}$ (see Eq. (\ref{AA})). It is thus straightforward to replace the approximated expression of the function
${\cal A}$ with its exact expression, as was done at fixed order.
All the remaining mass effects enter through the matching procedure in Eq.~(\ref{resfin}) and are treated at fixed order. They can be thus dealt with analogously to what is done in the fixed order calculation and described in Sec. \ref{sec:HNNLO}.

\subsection{Bottom quark loop}

\label{sec:bottom}

The implementation of mass effects outlined in the previous section is straightforward and is certainly appropriate when the heavy-quark mass is of the order of
the Higgs mass, as in the case of the top quark.
When the heavy-quark mass is much smaller than the Higgs mass, as in the case of the bottom quark,
the computation of the $\pT$ spectrum becomes a {\em three scale} problem, whose solution beyond fixed order
is by far non trivial.

The impact of the heavy-quark masses in the $\pT$ spectrum at NLO was shown in Fig.~\ref{fig:ptdistmtmb}. When only the top quark effect is considered, the behaviour of the $\pT$ spectrum in the small-$\pT$ region is very similar to the corresponding behaviour in the large-$\mt$ limit. In this region in fact the $\pT$ spectrum is divergent, and the singularity is driven by the
well known soft and collinear singularities of the relevant $H+3$ parton matrix element.
Since such singular behaviour is universal, the shape of the spectrum in this $\pT$ region does not depend on the inclusion
of the top-quark mass.

When the bottom-quark mass is included, the behaviour of the spectrum is rather different.
When $\pT\ltap \mb$ the behaviour is still driven by the singularities of the $H+3$ parton matrix element, but in the region $\mb\,\ltap\, \pT<\mH$ the shape of the spectrum is distorted.

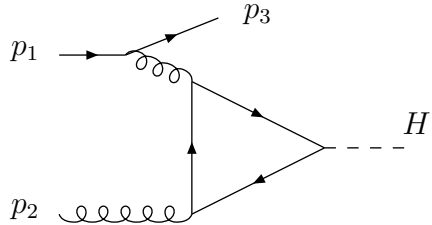
\begin{figure}[hbt]
\begin{center}
\SetScale{1}
\begin{picture}(120,100)(0,-90)
\Gluon(20,-70)(70,-70){-3}{5}
\ArrowLine(20,-10)(45,-10)
\Gluon(45,-10)(70,-20){3}{3}
\ArrowLine(45,-10)(80,4)
\ArrowLine(70,-70)(70,-20)
\ArrowLine(70,-20)(120,-45)
\ArrowLine(120,-45)(70,-70)
\DashLine(120,-45)(150,-45){5}
\put(2,-10){$p_1$}
\put(2,-70){$p_2$}
\put(90,4){$p_3$}
\put(150,-40){$H$}
\end{picture}  \\
\caption{\label{fig:qgchannel} \it Typical Feynman diagrams for the $qg\to Hq$ process.}
\end{center}
\end{figure}

In order to better understand what happens, in the following we examine the analytic behaviour
of the QCD matrix elements \cite{Ellis:1987xu}.
To make the discussion simpler let us consider the amplitude of the Higgs production in the $qg\rightarrow Hq$ channel. In this channel only one Feynman diagram contributes, which is shown in Fig. \ref{fig:qgchannel}: it consists in a triangular loop in which one of the
gluons is off shell and radiated from the incoming quark line.
In the small-$\pT$ region the singular behaviour is due to the collinear region, in which the gluon with momentum $p_1-p_3$ goes on shell.

\begin{figure}[htb]
\begin{center}
\includegraphics[scale=0.55]{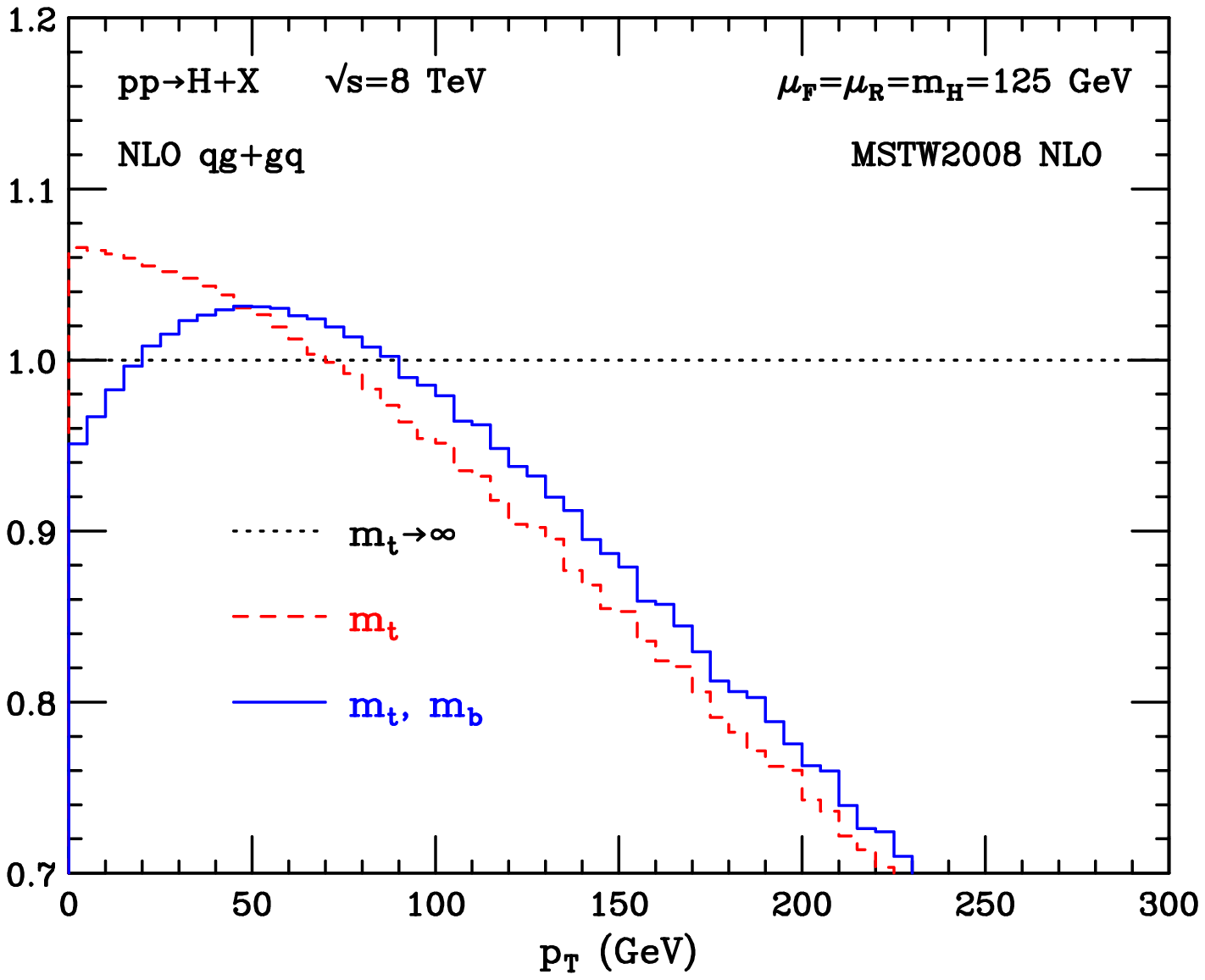}
\includegraphics[scale=0.55]{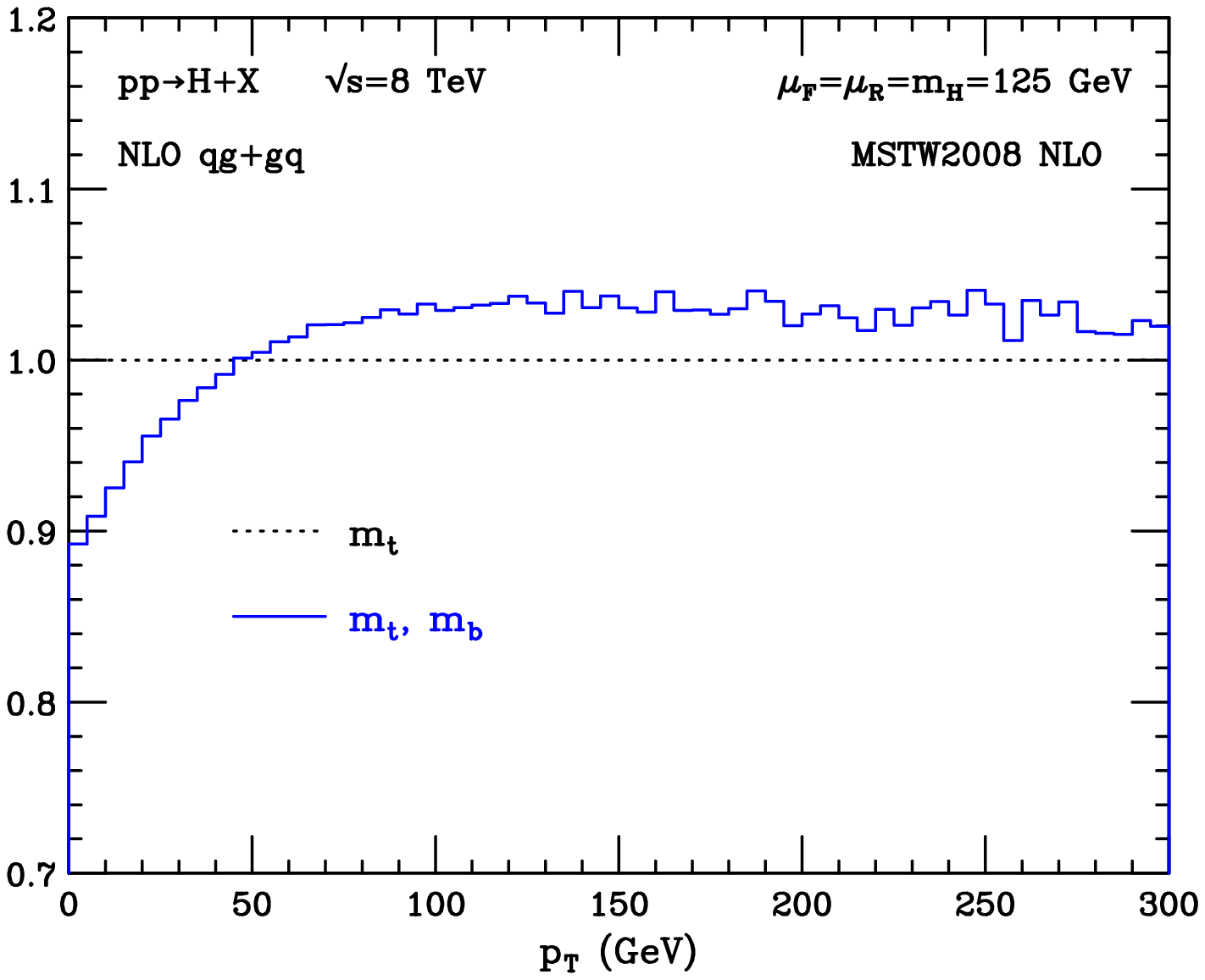}\\
\caption{Transverse momentum distribution for a SM Higgs with $m_H=125$ GeV computed at NLO in the $qg+gq$ channel. Left: result normalized to the large-$\mt$ approximation. Right: normalized to the $m_t$-dependent result.}
\label{fig:HResgq}
\end{center}
\end{figure}

In Fig. \ref{fig:HResgq} we plot the $p_T$ spectrum in this channel normalized to the corresponding result in the large-$m_t$ limit. We see that the qualitative behaviour is the same observed in Fig. \ref{fig:ptdistmtmb}: the behaviour of the spectrum is distorted in the small and intermediate
$\pT$ region by the presence of the bottom mass.

%%%%%%%%%%%%%%%%%%%%%%%%%%%%%%%%%%%%%%%%%%%%%%%%%
%%%%%%%%%%%%%%%%%%%%%%%%%%%%%%%%%%%%%%%%%%%%%%%%%

This behaviour is somewhat against intuition: in the region $\pT\ll \mH$ we could expect the cross section to factorize naively as in the large-$\mt$ limit, and the structure of the triangular loop
to affect only the overall normalization. This would lead to
a similar behaviour of the exact and approximated $\pT$ spectra, and
thus to a relatively flat behaviour of the histogram in Fig.~\ref{fig:HResgq}.
As we will see below, such behaviour is disrupted by the effect of the bottom mass.

The matrix element squared for the process $q(p_1)+g(p_2)\rightarrow q(p_3)+H$ is given by \cite{Ellis:1987xu}
\begin{equation}
\lvert\mathcal{M}_{qg\to qH}(s,t,u)\rvert^2=\alpha_W\as^3\, C_FC_A \frac{u^2+s^2}{(-t)\,m_W^2}\frac{m_H^4}{(u+s)^2}\lvert A_5(t,s,u)\rvert^2,
\label{qgH}
\end{equation}
where $\alpha_W$ is the EW coupling, $m_W$ is the $W$ mass, and $s,t$ and $u$ are the Mandelstam invariants
\begin{equation}
s=(p_1+p_2)^2, \ \, t=(p_1-p_3)^2, \ \, u=(p_2-p_3)^2,
\label{kinematic}
\end{equation}
satisfying
\begin{equation}
s+u+t=m_H^2, \ \ \, ut=sp_T^2.
\label{kinematic2}
\end{equation}
The physical region of these invariants is $u\leq 0, \ \, t\leq 0$ and $s\geq m_H^2$.
The function $A_5(t,s,u)$ is given by
\begin{align}
A_5(t,s,u)&=\sum_{f}\frac{m_f^2}{m_H^2}\Bigl[4+\frac{4t}{u+s}\left[W_1^f(t)-W_1^f(m_H^2)\right]\\
&+\left[1-\frac{4m_f^2}{u+s}\right]\left[W_2^f(t)-W_2^f(m_H^2)\right]\Bigr],
\label{A5}
\end{align}
where the summation is over the heavy-quark running in the triangular loop.
The explicit expressions for the functions $W_1^f$ and $W_2^f$ are given in Ref. \cite{Ellis:1987xu}. In the physical region ($t\leq 0$) they read
\begin{equation}
W_1^f(t)=2(1+4m_f^2/(-t))^{1/2} {\rm arcsinh}\left(\sqrt{-t}/(2m_f)\right)
\hskip 1cm 
W_2^f(t)=4{\rm arcsinh}^2\left(\sqrt{-t}/(2m_f)\right)\, .
\end{equation}
In the limit $|t|\ll 4m_f^2$ we have
\begin{equation}
W_1^f(t)\to 2~~~~~W_2^f(t)\to 0
\end{equation}
and the amplitude $A_5(t,s,u)$ reduces to
\begin{equation}
A_5(t,s,u)\to A_1(m_H^2)=\sum_{f}\frac{m_f^2}{m_H^2}\left[4-W_2^f(m_H^2)\left(1-\frac{4m_f^2}{m_H^2}\right)\right]\, ,
\label{A1}
\end{equation}
where $A_1(m_H^2)$ is the Born $gg\rightarrow H$ amplitude \cite{Georgi:1977gs}.
In this limit the matrix element squared $\lvert\mathcal{M}_{qg\to qH}(s,t,u)\rvert^2$ can thus
be written as
\begin{equation}
\label{naivecol}
\lvert\mathcal{M}_{qg\to qH}(s,t,u)\rvert^2\to \alpha_W\as^3 C_A\f{\mH^4}{(-t)m^2_W\, z}
{\hat P}_{gq}(z) \lvert A_1(\mH^2)\rvert^2\, ,
\end{equation}
where 
\begin{equation}
{\hat P}_{gq}(z)=C_F\, \f{1+(1-z)^2}{z}
\end{equation}
and $z=\mH^2/s$.

As it is mentioned above, we are interested in the $p_T\sim m_b$ region. In this case one still can use the $|t|\to 0$ limit for the terms coming from the top quark contribution, while for the terms involving the bottom quark this approximation is not justified.  The bottom quark contribution to $A_5(t,s,u)$ in the region of interest can be written as
\begin{equation}
A_{5b}(t,s,u)=\frac{m_b^2}{m_H^2}\left[4-\left(W_2^b(m_H^2)-W_2^b(t)\right)\left(1-\frac{4m_b^2}{m_H^2}\right)\right],
\label{A5b}
\end{equation}
which differs from Eq. (\ref{A1}) by the presence of the $W_2^b(t)$ contribution.
In the region in which $\lvert t\rvert\sim 4m_b^2$ the term proportional to $W_2(t)$ in Eq. (\ref{A5b}) cannot be neglected. This means that naive collinear factorization, which would lead us to recover the Born result in Eq. (\ref{A1}), and thus Eq.~(\ref{naivecol}), is not a good approximation here. Furthermore, $W_2(t)$ is an increasing function of $-t$, and hence, of $p_T$, thus explaining the steep behaviour in Fig. \ref{fig:HResgq}.

The analytic behaviour in the $gg$ channel is more complicated but the physical picture remains the
same: when the transverse momentum of the Higgs boson is much smaller than its mass, but of the order of the bottom quark mass, the naive factorization valid in the large-$\mt$ limit does not apply.
This implies that the resummed calculation of the Higgs $\pT$ spectrum, which is based on the assumption that only the two scales $\pT$ and $\mH$ are relevant, cannot be straightforwardly extended
to include the bottom quark. The resummation formalism is in fact based on the possibility to
factorize and resum the emission of multiple soft and collinear partons from the underlying
Born subprocess.
%%%%%%  ADDED IN REVISED VERSION %%%%%%
In other words, the presence of non factorizable contributions at $p_T$ of order $m_b$
and larger not only spoils soft and collinear factorization beyond this region, but introduces
an effective scale of of the order of $m_b$ which acts as a cut off in the
integration of the soft and collinear spectrum, effectively replacing
the natural scale $m_H$ in the logarithmically enhanced terms.
%%%%%%%%%%%%%%%%%%%%%%%%%%%%%%%%%%%%%%%

In the following we propose a simple solution to this problem, which is based on the following
observation.
The dominant contribution to the transverse momentum cross section is given by the top quark, and, in this case,
the resummation procedure is fully justified.
The bottom-quark contribution, however, is important to obtain a reliable prediction for the transverse momentum spectrum.
Since it introduces an additional scale that complicates the resummation of the large logarithmic terms,
it would be better to treat it at fixed order.
On the other hand, although small with respect to the top-quark contribution, the bottom contribution is still divergent as $\pT\to 0$. We thus would like the resummation to be effective only in the region $\pT\ltap \mb$, where it is indeed needed.
The resummation formalism of Ref.~\cite{Bozzi:2005wk} does
allow to achieve this goal, since it introduces an additional scale, the resummation scale (see Eq.~(\ref{logdef})), which actually controls the transverse momentum region up to which resummation is effective.

We can thus proceed as follows. We split the calculation in two parts. The first part includes only the top-quark contribution, and can thus be treated up to NLL+NLO as discussed in Sec.~\ref{sec:resu}, with a resummation scale $Q_1\sim \mH$.
The second part includes the bottom-quark contribution and the top-bottom interference, and we treat it separately, by choosing a resummation scale $Q_2\sim \mb$. 
This procedure can be implemented through the
following replacement in the resummed form factor of Eq.~(\ref{wtilde})
\begin{equation}
{\cal W}^{(N_1,N_2)}\longrightarrow  {\cal W}_{\rm top}^{(N_1,N_2)}+{\cal W}_{\rm bot}^{(N_1,N_2)}
\end{equation}
where
\begin{align}
{\cal W}_{\rm top}^{(N_1,N_2)}(b)&=
\sigma_{LO}(\mt){\cal H}^{(N_1,N_2)}(m_H^2/Q_1^2;\mt)\,
\exp\{{\cal G}^{(N_1,N_2)}({\widetilde L}_{Q_1};m_H^2/Q_1^2)\}\\
{\cal W}_{\rm bot}^{(N_1,N_2)}(b)&=
\Big[\sigma_{LO}(\mt,\mb){\cal H}^{(N_1,N_2)}(m_H^2/Q_2^2;\mt,\mb)
-\sigma_{LO}(\mt){\cal H}^{(N_1,N_2)}(m_H^2/Q_2^2;\mt)\Big]\nn\\
&\times\exp\{{\cal G}^{(N_1,N_2)}({\widetilde L}_{Q_2};m_H^2/Q_2^2)\}\, ,
\end{align}
and the dependence on $\as$ and $\mH$ has been understood.
We can finally add the NNLL+NNLO terms for the top quark contribution
${\cal W}_{\rm top}^{(N_1,N_2)}(b)$
in the large-$\mt$ limit by including the coefficients ${\cal H}^{(2)}$ \cite{Catani:2011kr}
and the NNLL expression for the form factor $\exp\{\cal{G}\}$.

We note that by using this procedure the standard resummed result is automatically
recovered by setting $Q_2=Q_1$. We also note that, thanks to the
unitarity constraint fulfilled by our formalism, the exact heavy-quark mass dependence up to NLO is recovered when integrating over $\pT$.
Since the inclusive cross section is known to contain logarithmic terms of
the form $\ln \mH/\mb$, such unitarity constraint implies that these logarithmic terms are correctly recovered (and not resummed) in the inclusive cross
section.

%%%%%%%%%%
\subsection{Numerical results}
\label{sec:results}

\begin{figure}[htb]
\begin{center}
\includegraphics[scale=0.8]{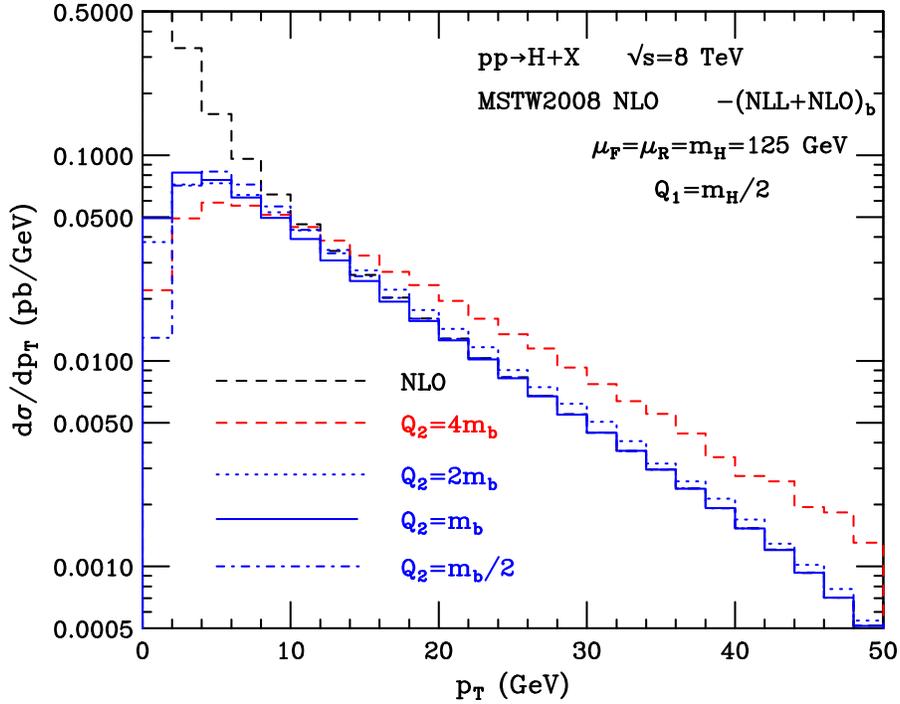}
\caption{Bottom quark contributions to the transverse momentum spectrum of the Higgs boson at NLO and NLL+NLO for different choices of the scale $Q_2$.}
\label{fig:tb}
\end{center}
\end{figure}

We start the presentation of our results by considering the bottom quark contribution\footnote{Here and in the following, when we mention the ``bottom quark contribution'' we always refer to all the perturbative contributions involving a bottom-quark loop, i.e., purely bottom contributions and top-bottom interferences.}.
Such contribution is dominated by the (negative) top-bottom interference.
In the small-$p_T$ region we resum the logarithmically enhanced terms up to NLL
and do the matching to the bottom-quark contribution at NLO.
Although these perturbative contributions are non physical,
the way they are treated in the full calculation is essential.
Since the scale up to which this resummation makes sense is lowered by the non factorizable contributions discussed in Sec.~\ref{sec:bottom}, we choose the
corresponding resummation scale $Q_2$ of the order of the bottom mass.
In Fig.~\ref{fig:tb} our NLL+NLO result for $Q_2=\mb/2,\mb,2\mb,4\mb$ is compared
to the fixed order NLO result. We see that the NLO result is divergent as $\pT\to 0$.
We also notice that, as expected, increasing $Q_2$ makes the resummed $\pT$ spectrum harder.
We see that, for $Q_2=\mb/2,\mb,2\mb$ there is a rather good agreement of the resummed and fixed
order results at $\pT\gtap 10$ GeV. On the contrary, the NLL+NLO result for $Q_2=4\mb$ does not
agree with the fixed order result in this region.
Since we want our resummation to be effective only in the very small $p_T$ region, we
choose $Q_2=\mb$ as our central scale choice, and proceed by performing the full NLL+NLO calculation. 
Thanks to the unitarity constraint discussed in the previous section, the integral of our NLL+NLO result coincides
with the inclusive NLO cross section. Since for $\mH=125$ GeV $\sigma_{NLO}(\mt,\mb)/\sigma_{NLO}(\mt\to \infty)\sim 0.988$ (see Table~1), the inclusion of heavy quark masses affects the shape of the spectrum, by leaving the normalization of the transverse momentum cross section essentially unchanged.

%%%%%%%%%%%%%%%%%%%%%%%%%%%%%%%%%%%%%%%%%%%%%%%%%
%%%%%%%%%%%%%%%%%%%%%%%%%%%%%%%%%%%%%%%%%%%%%%%%%

\begin{figure}[htb]
\begin{center}
\includegraphics[scale=0.8]{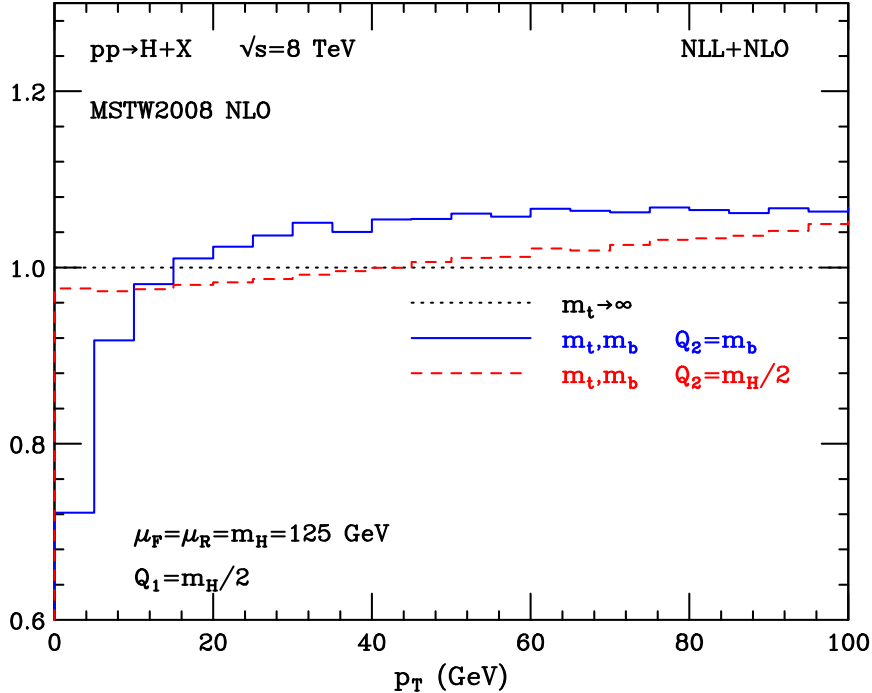}
\caption{Transverse momentum spectra at NLL+NLO for $Q_2=\mb$ and $Q_2=\mH/2$ normalized to the result in the large-$\mt$ limit.}
\label{fig:compratio}
\end{center}
\end{figure}

In Fig.~\ref{fig:compratio} we present our resummed spectrum computed with $Q_2=\mb$ as discussed above, normalized to the corresponding result in the large-$\mt$ limit, and we compare it with the corresponding result obtained ignoring the problem with the bottom quark contribution, and setting $Q_2=Q_1=\mH/2$.
We see that the impact in the shape is significant. The choice $Q_2=\mb$ makes the resummed
$\pT$ spectrum harder. This is not unexpected: since the resummed bottom-quark contribution is negative,
choosing a smaller resummation scale for it reduces the cross section at
small $\pT$ and increases it at intermediate and large values of $\pT$.
%%%% ADDED IN REVISED VERSION %%%%%
More precisely, the choice $Q_2=\mb$ increases the cross section by about $1\%$ ($5\%$) at $\pT=100$ GeV ($\pT=40$ GeV), and decreases the cross section
by about $25\%$ in the first bin.
%%%%%%%%%%%%%%%%

%%%%%%%%%%%%%%%%%%%%%%%%%%%%%%%%%%%%%%%%%%%%%%%%%
%%%%%%%%%%%%%%%%%%%%%%%%%%%%%%%%%%%%%%%%%%%%%%%%%

The heavy-quark mass effects in the resummed $\pT$ spectrum
were first implemented up to NLL+NLO in Ref.~\cite{Mantler:2012bj}. 
In the latter paper the top and bottom quarks are treated on the same
footing and the resummed calculation for the $\pT$ spectrum
corresponds to the case in which $Q_2=Q_1=\mH/2$. 
By comparing the dashed histogram in Fig.~\ref{fig:compratio} with the corresponding curve in Fig.~6 of Ref.~\cite{Mantler:2012bj} we find relatively good agreement, despite the fact that Ref.~\cite{Mantler:2012bj} uses $\mu_F=\mu_R=\mH/2$.
The good agreement is confirmed when we adopt the same scale choice.
%%%%% ADDED IN REVISED VERSION %%%%%
However, as discussed in Sect.~\ref{sec:bottom},
the choice $Q_2=Q_1$ corresponds to ignore
the factorization breaking in the bottom-quark contribution,
and, in our opinion, it is not advisable.
%%%%%%%%%%%%%%%%%%%%%%%%%%%%%%%%%%%%%

The resummation of the logarithmically enhanced terms in the $\pT$ spectrum
is effectively
performed by Monte Carlo event generators. The method of matching
NLO computations to parton shower simulations,
implemented in {\tt MC@NLO} \cite{Frixione:2002ik} and {\tt POWHEG} \cite{Frixione:2007vw},
allows the user to achieve an accuracy which is roughly comparable to the
accuracy of our resummed NLL+NLO calculation.
Such Monte Carlo
generators have traditionally used the large-$\mt$ approximation in
their implementations of Higgs production through gluon fusion.
Recently, the exact top- and bottom-mass dependence has been implemented both
in {\tt MC@NLO} \cite{Frixionetalk} and {\tt POWHEG} \cite{Bagnaschi:2011tu}.
A comparison of the relative effect of the exact top-mass dependence
with respect to the result in the large-$\mt$ limit shows
a good agreement between the two generators. The inclusion of the
bottom-quark mass leads instead to relatively large differences \cite{Frixionetalk}.

Comparing our results with those of \cite{Frixionetalk} we
find that, in the case of $Q_1=Q_2$ the quantitative
impact of the bottom quark on the shape of the $\pT$ spectrum
is very similar to what found with {\tt MC@NLO},
while {\tt POWHEG} somewhat amplifies the effect of the bottom quark.
%
%%%%% ADDED IN REVISED VERSION %%%%%%%
This is not unexpected: the matching procedure implemented in {\tt MC@NLO}
carries many similarities to the one adopted in {\tt HRes}, the difference being that, while
in {\tt HRes} the resummation is carried out analytically (see Sec.~\ref{sec:resu}), in {\tt MC@NLO} it
is performed through the parton shower.
On the contrary, {\tt POWHEG} works rather differently: since it
exponentiates the full real emission matrix element,
the bottom-quark contribution is expected to affect the spectrum in a different way.
Nonetheless,
the arguments of Sect.~\ref{sec:bottom} apply not only to analytical resummation,
but also to Monte Carlo simulations. Since both {\tt MC@NLO} and {\tt POWHEG}
treat the top and bottom contributions on the same footing,
we do not regard the ensuing results as theoretically motivated.
With our default choice of $Q_2=\mb$ the shape of the spectrum is (accidentally)
more similar to the {\tt POWHEG} result, though in our case
the effects of the bottom quark
are confined to smaller values of $\pT$.
%%%% END %%%%
%With our default choice of $Q_2=\mb$ the shape is somewhat
%more similar to the {\tt POWHEG} result, though in our case
%the effects of the bottom quark
%are confined to smaller values of $\pT$.

%%%%%%%%%%%%%%%%%%%%%%%%%%%%%%%%%%%%%%%%%%%%%%%%%
\begin{figure}[htb]
\begin{center}
\includegraphics[scale=0.55]{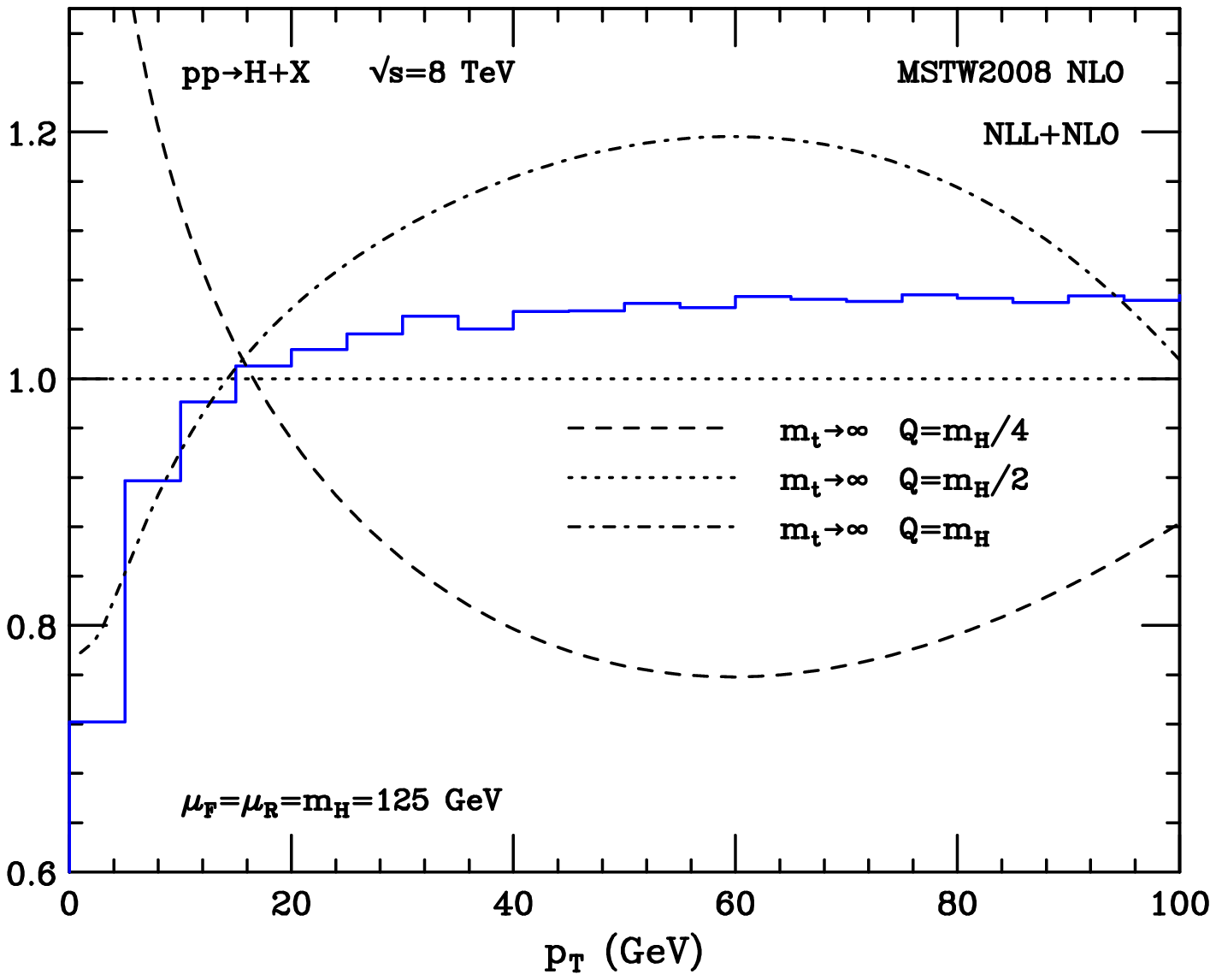}
\includegraphics[scale=0.55]{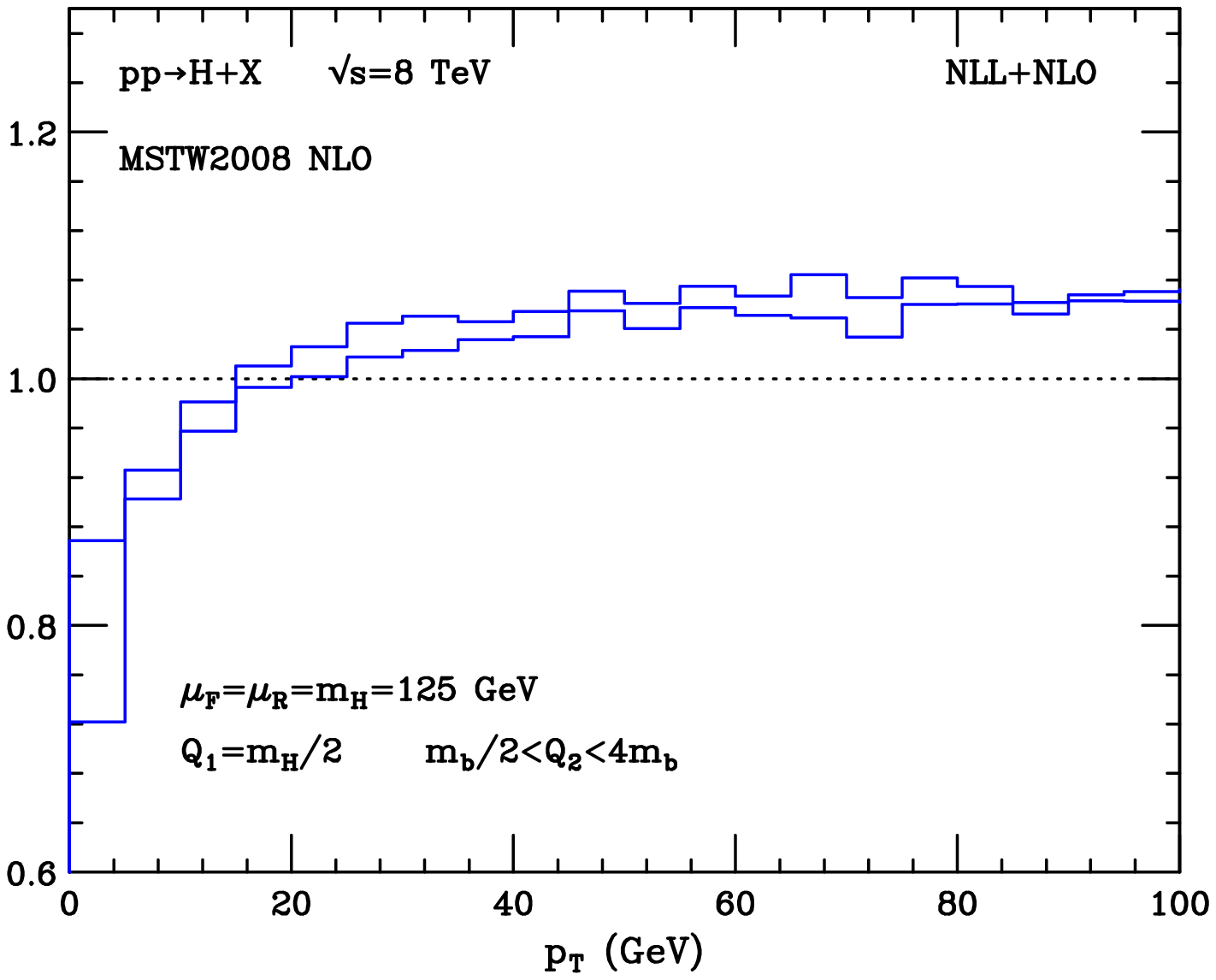}
\caption{Left: transverse momentum spectrum at NLL+NLO with full dependence on heavy quark masses ($Q_2=\mb$) normalized to the result in the large-$\mt$ limit (solid histogram). The result is compared to the NLL+NLO results in the large-$\mt$ limit obtained with $Q=\mH,\mH/4$. Right: transverse momentum spectra at NLL+NLO for $\mb/2<Q_2<4\mb$ normalized to the result in the large-$\mt$ limit.}
\label{fig:ratio}
\end{center}
\end{figure}
%%%%%%%%%%%%%%%%%%%%%%%%%%%%%%%%%%%%%%%%%%%%%%%%%

In order to assess the relevance of heavy-quark mass effects at
%this perturbative order,
NLL+NLO,
it is important to
compare their quantitative impact to the uncertainties affecting the resummed $\pT$ spectrum computed
in the large-$\mt$ limit.
At NLL+NLO, it is known that perturbative uncertainties are relatively large.
While variations of the renormalization and factorization scales affect both the shape and the normalization
of the $\pT$ cross section, the choice of the resummation scale $Q$ affects only the shape of the spectrum.
In particular, as discussed above, increasing (decreasing) $Q$ makes the spectrum harder (softer).
In Fig.~\ref{fig:ratio} (left) we present our resummed spectrum at NLL+NLO with inclusion of the heavy-quark masses as in Fig.~\ref{fig:compratio}, and compare it with the spectrum computed in the large-$\mt$ limit for $Q=\mH/4,\mH$ with the numerical program {\tt HqT}. We see that, as anticipated, the effect of resummation scale variations is large, well beyond the effect of heavy-quark masses for $\pT\gtap 20$ GeV. In the region $\pT\ltap 20$ GeV instead, the effect of the bottom-quark mass and of resummation scale variation are comparable.
%%% ADDED IN REVISED VERSION %%%
In Fig.~\ref{fig:ratio} (right) we study the impact of $Q_2$ variations on the $\pT$ spectrum. Since we have chosen $Q_2=\mb$ as central value of the scale, the standard practice would suggest to vary $Q_2$ between $\mb/2$ and $2\mb$ to estimate the theoretical uncertainty. However, we should not forget that in the region $\mb\ltap \pT\ltap \mH$ there are potentially large logarithmic contributions beyond ${\cal O}(\as^3)$ that in our approach are treated essentially at fixed order. To be conservative, we thus extend the upper limit of $Q_2$ to $4\mb$. We see that the impact of $Q_2$ variations is moderate, at the level of $\pm 1-2\%$, except in the first bin, where it reaches about $\pm 10\%$.  This is also the region of transverse momenta where the effect of the bottom quark is more significant.
%In Fig.~\ref{fig:ratio} (right) we present our resummed spectrum computed with $Q_2=\mb/2,\mb,2\mb$.
%We see that the effects of $Q_2$ variations around $\mb$ is relatively small, and leaves the shape of the
%distribution rather stable.
%

\begin{figure}[htb]
\begin{center}
\includegraphics[scale=0.55]{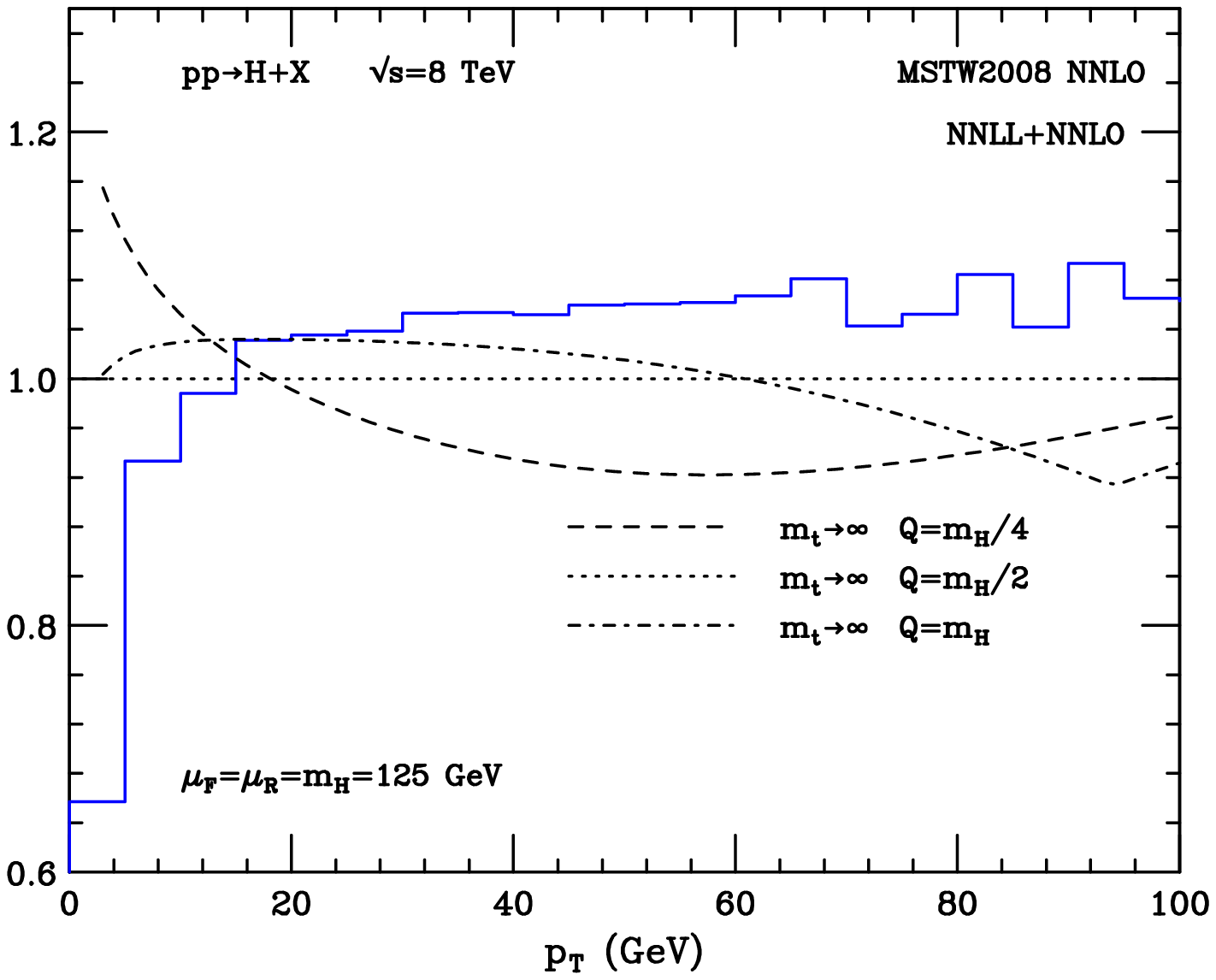}
\includegraphics[scale=0.55]{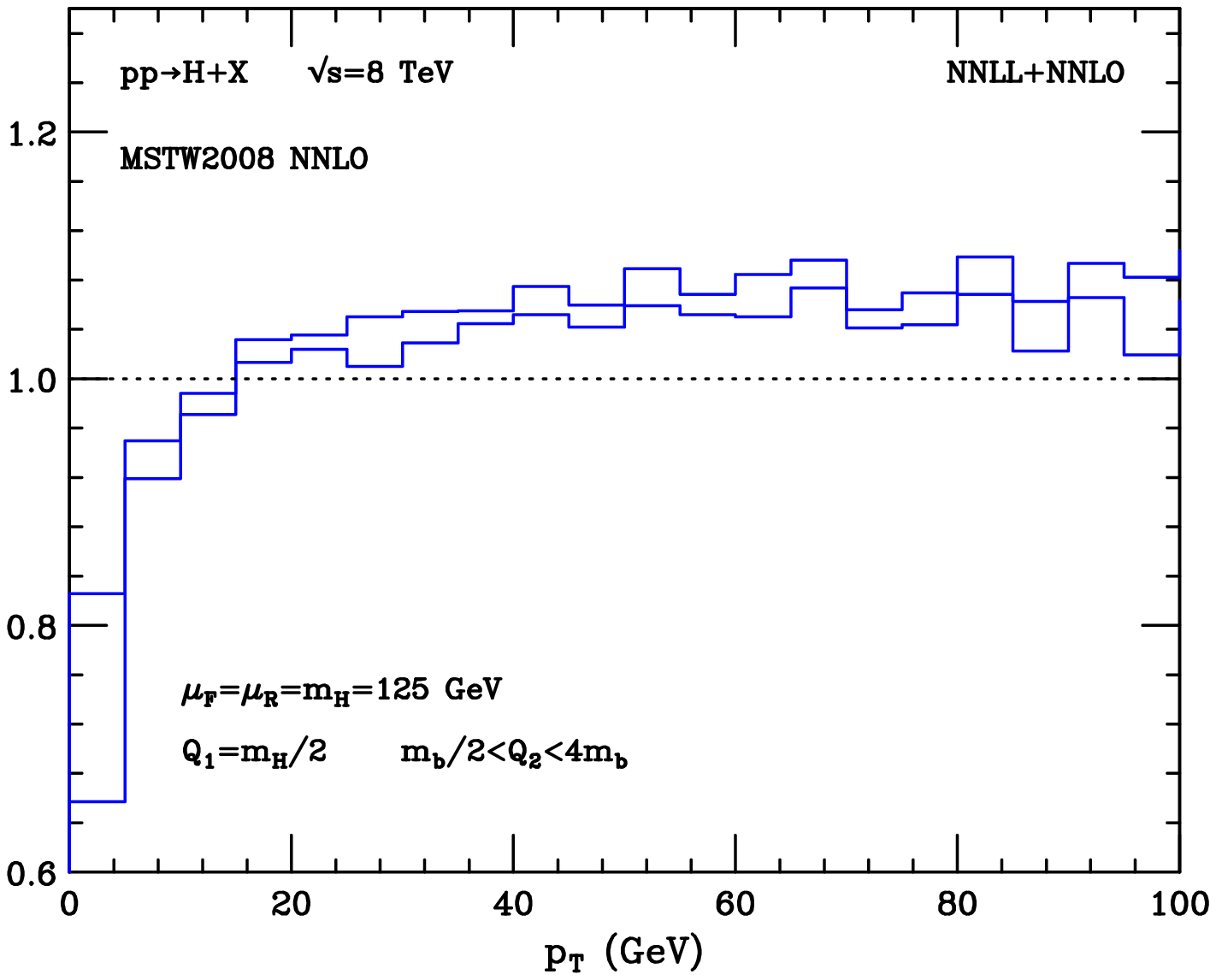}
\caption{The same as in Fig.~\ref{fig:ratio} but at NNLL+NNLO.}
\label{fig:rationnll}
\end{center}
\end{figure}

We now move to consider the NNLL+NNLO results.
Since for $\mH=125$ GeV we have\footnote{We remind the reader (see Sec.~\ref{sec:HNNLO}) that the ${\cal O}(\as^4)$ terms in our calculation are rescaled with the exact $\mt$ dependent Born cross section.} $\sigma_{NNLO}(\mt,\mb)/\sigma_{NNLO}(\mt\to \infty)\sim 1.007$, the inclusion of heavy quark masses, as happens at NLL+NLO, affects the shape of the spectrum, but leaves the normalization of the transverse momentum cross section essentially unchanged.
In Fig.~\ref{fig:rationnll} (left) the NNLL+NNLO spectrum normalized to the NNLL+NNLO result in the large-$\mt$ limit is presented, and compared to the large-$\mt$ limit results for $Q=\mH/4$ and $Q=\mH$. Comparing with Fig.~\ref{fig:ratio} (left) we see that the impact of heavy-quark mass effects in the NNLL+NNLO result is similar to what observed at NLL+NLO. This should somewhat be expected, since the NNLL+NNLO terms we are adding are evaluated in the large-$\mt$ limit.
We notice that, as is known \cite{Bozzi:2005wk}, the effect of resummation scale variations at this order is
much smaller and we conclude that
mass effects in the low-$\pT$ region are even more important.
%%% ADDED IN REVISED VERSION %%%
In Fig.~\ref{fig:rationnll} (right) we study the impact of $Q_2$ variations at this order.
As in Fig.~\ref{fig:ratio}, we vary $Q_2$ in the range $\mb/2<Q_2<4\mb$.
We see that the effects is moderate and similar to what found at NLL+NLO.
%%%%%%

%%%%%%%%%%%%%%%%%%%%%%%%%%%%%%%%%%%%%%%%%%%%%%%%%
%%%%%%%%%%%%%%%%%%%%%%%%%%%%%%%%%%%%%%%%%%%%%%%%%

We finally add few comments on the uncertainties affecting the {\em shape} of the
resummed $\pT$ spectrum at NNLL+NNLO. Such uncertainties were previously studied in Ref.~\cite{deFlorian:2011xf},
within the large-$\mt$ approximation, by using the {\tt HqT} numerical program.
It was shown (see Fig.~4 of Ref.~\cite{deFlorian:2011xf}) that the combined variation of resummation, renormalization and factorization scale leads to effects up to ${\cal O}(\pm 5\%)$ on the normalized spectrum $1/\sigma \times d\sigma/d\pT$.
Non-perturbative effects (see Ref.~\cite{Bozzi:2005wk} and references therein)
are expected to significantly affect the $\pT$ distribution only in the region $\pT\ltap 10$ GeV, and, in particular, to make the spectrum (slightly) harder. 
Compared to the above effects, the impact of heavy quark-masses discussed in this paper
is certainly relevant, since it leads
to a distortion of the spectrum at least of the same order,
and definitely larger at small $\pT$.
%%%%%%%% ADDED IN REVISED VERSION %%%%%%%%
From the previous discussion we conclude that
the estimated uncertainty from the treatment of the bottom quark
on the shape of the spectrum is typically small, at the $\pm 1-2\%$ level, but it increases to the $\pm 10\%$ level at very small $\pT$ ($\pT\ltap \mb$).

\section{Conclusions}
\label{sec:summa}

In this paper we have considered heavy-quark mass effects in Higgs boson production through gluon fusion at the LHC. We have extended previous computations
of the fully exclusive $gg\to H$ cross section and of the resummed $\pT$ spectrum
by implementing the exact top- and bottom-mass dependence up to ${\cal O}(\as^3)$.

The implementation of the top-mass dependence does not lead to substantial complications.
By contrast, since $\mb\ll \mH$, the inclusion of the exact bottom-mass dependence in the resummed $\pT$ spectrum implies the solution of a non-trivial three-scale problem.
We have studied the analytical behaviour of the relevant QCD matrix elements, showing that,
when the bottom-quark contribution is considered, naive factorization is valid only in a limited region of the phase space, i.e. when $\pT\ltap 2\mb$.
We have provided a simple solution to this issue by controlling the resummed bottom-quark contribution
through an additional resummation scale $Q_2$, which was chosen of the order of the
bottom mass $\mb$. We have shown that this
solution has a clear advantage: it limits the impact of the resummation
to the region where it is really needed, i.e. $\pT\ltap 2\mb$, and our resummed result for the bottom quark contributions smoothly merges with the fixed order NLO result at $\pT\gtap 2\mb$, where the resummation is not anymore justified (see Fig.~\ref{fig:tb}).

We have studied the impact of mass effects on the NLL+NLO calculation, by showing that the effect of heavy-quark masses is significant, although
at this order large uncertainties affect the resummed $\pT$ spectrum.
When going to NNLL+NNLO, where the perturbative uncertainties are much smaller, the impact of heavy-quark masses
is even more relevant, since it significantly distorts the spectrum in the low-$\pT$ region.
Our calculations are implemented in updated versions of the {\tt HNNLO} and {\tt HRes} numerical programs.

\noindent {\bf Acknowledgements.}
We would like to thank Emanuele Bagnaschi for useful correspondence, Stefano Catani, Lance Dixon and Daniel de Florian for helpful discussions and comments on the manuscript.
This research was supported in part by
the Research Executive Agency (REA) of the European Union under the Grant Agreement number PITN-GA-2010-264564 ({\it LHCPhenoNet}).

\bibliographystyle{atlasnote}
\bibliography{masseffects}

\end{document}